\documentclass[aps, pra, citeautoscript, 10pt, notitlepage, superscriptaddress, twocolumn]{revtex4-2}

\usepackage[margin=1.54cm]{geometry}
\usepackage{algorithm}
\usepackage{algpseudocode}
\usepackage{graphicx}
\usepackage{amsmath,mathtools}
\usepackage[hidelinks]{hyperref}
\usepackage{amssymb}
\usepackage{siunitx}
\usepackage{color}
\usepackage[ngerman,english]{babel}
\usepackage{physics}
\usepackage{bm}
\usepackage{natbib}
\usepackage{booktabs}
\usepackage{float}
\begin{document}

\title{Role of Spatial Coherence in Diffractive Optical Neural Networks}

\author{Matthew J. Filipovich}
\email{matthew.filipovich@physics.ox.ac.uk}
\author{Aleksei Malyshev}
\affiliation{Clarendon Laboratory, University of Oxford, Parks Road, Oxford, United Kingdom}
\author{A. I. Lvovsky}
\affiliation{Clarendon Laboratory, University of Oxford, Parks Road, Oxford, United Kingdom}
\affiliation{Lumai, Wood Centre for Innovation, Quarry Road, Headington, Oxford, United Kingdom}



\begin{abstract} 
Diffractive optical neural networks (DONNs) have emerged as a promising optical hardware platform for ultra-fast and energy-efficient signal processing for machine learning tasks, particularly in computer vision. Previous experimental demonstrations of DONNs have only been performed using coherent light. However, many real-world DONN applications require consideration of the spatial coherence properties of the optical signals. 
Here, we study the role of spatial coherence in DONN operation and performance.
We propose a numerical approach to efficiently simulate DONNs under incoherent and partially coherent input illumination and discuss the corresponding computational complexity.
As a demonstration, we train and evaluate simulated DONNs on the MNIST dataset of handwritten digits to process light with varying spatial coherence.
\end{abstract}

\maketitle 

\section{Introduction}

Machine learning has led to breakthroughs in many fields of technology including computer vision, natural language processing, and drug discovery~\cite{Goodfellow-et-al-2016, bochkovskiy_yolov4_2020, brown_language_2020, senior_improved_2020}. Currently, machine learning models are executed using specialized electronic hardware, such as graphics processing units (GPUs) and tensor processing units (TPUs), which harness immense processing power and data parallelism. However, the growing compute requirements from advanced deep learning models are far outpacing hardware improvements anticipated by Moore's law scaling~\cite{dario_amodei_ai_2018}. Consequently, progress in machine learning using current digital hardware will soon become technically and economically unsustainable, while also producing a significant environmental impact~\cite{thompson_computational_2022, mehonic_brain-inspired_2022}.

Given the constraints imposed by digital electronics, optics has gained recognition as a promising platform for machine learning applications with low latency, high bandwidth, and low energy consumption~\cite{shastri_photonics_2021, vuong_what_2023}.
Several optical implementations of neural networks have recently been demonstrated using both free-space optics~\cite{wang_image_2023, spall_fully_2020,miscuglio_massively_2020, zuo_all-optical_2019} and integrated photonics~\cite{zhang_silicon_2022,ashtiani_-chip_2022, feldmann_all-optical_2019, tait_neuromorphic_2017}.
Computational speeds of trillions of operations per second have been achieved using optical processing hardware~\cite{xu_11_2021, feldmann_parallel_2021}, and optical neural networks using less than one photon per multiplication have been experimentally realized~\cite{ma_quantum-noise-limited_2023, wang_optical_2022}.
Additionally, optical architectures for implementing \textit{in situ} training of neural networks have been demonstrated~\cite{spall_training_2023, pai_experimentally_2023, filipovich_silicon_2022,spall_hybrid_2022, zhou_situ_2020,hughes_training_2018}.

\begin{figure*}
    \centering
    \includegraphics[width=0.75\linewidth]{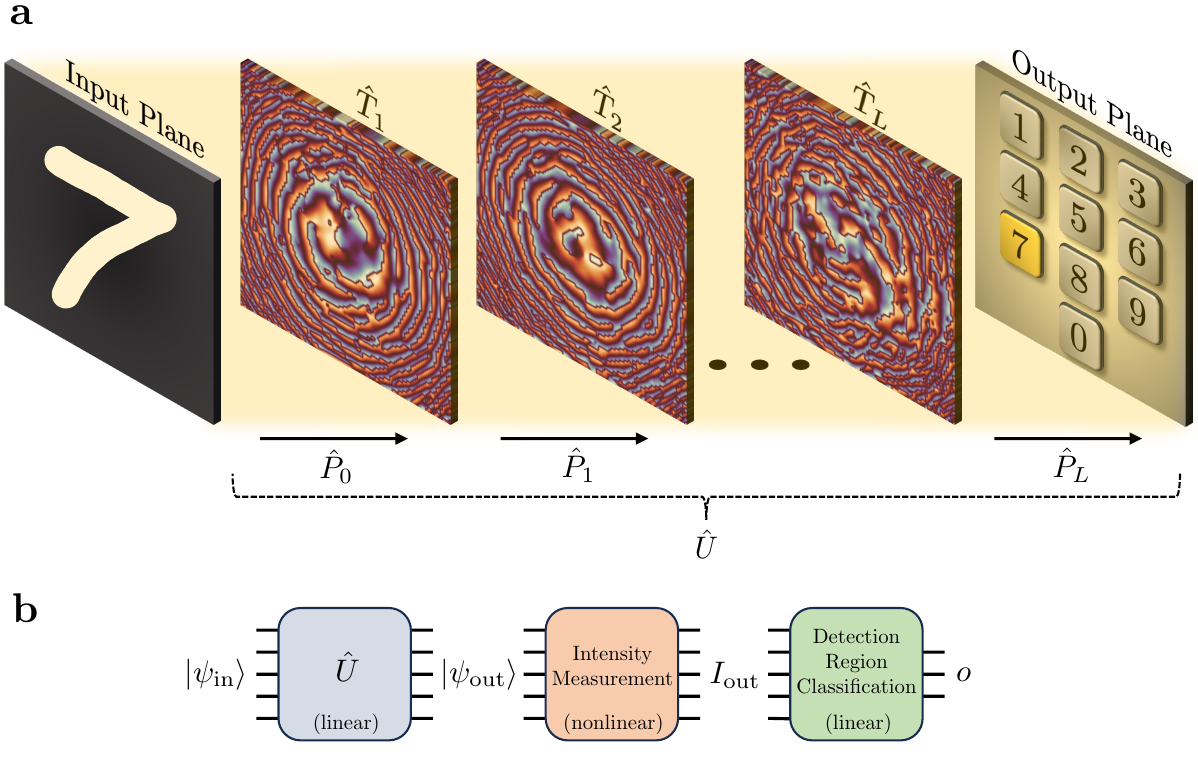}
    \caption{Diffractive optical neural network (DONN) architecture. \textbf{a}~Illustration of a DONN trained to identify handwritten digits. The DONN is comprised of $L$ diffractive layers that modulate the optical field as it propagates through the system.
    The output plane encompasses ten detection regions, which are each associated with a unique digit, and the predicted output corresponds to the region with the highest optical intensity.
    The transmission and propagation operators at the $l$-th layer are denoted by $\hat{T}_{l}$ and $\hat{P}_{l}$, respectively.   
    \textbf{b}~Summary of the transformations applied to the input field $\ket{\psi_\mathrm{in}}$ in a DONN. 
    The linear evolution operator $\hat{U}$ maps the input field onto the output field. The intensity $I_\mathrm{out}$ at the output plane is measured, and the output of the DONN, represented by $o$, corresponds to the total intensity incident on each detection region.
    }
    \label{fig:diffractive_network_schematic}
\end{figure*}

Diffractive optical neural networks (DONNs) are specialized hardware architectures that harness diffraction effects to process optical signals in free space~\cite{lin_all-optical_2018,mengu_analysis_2020}.
DONNs are generally composed of several successive modulation surfaces, denoted as diffractive layers, that modify the phase and/or amplitude of the incident optical signals through light-matter interactions, as shown in Fig.~\ref{fig:diffractive_network_schematic}a. The diffractive layers contain discrete pixels, each with an independent complex-valued transmittance coefficient. The output of the DONN corresponds to the total  intensity of the optical field incident on designated detection regions in the output plane.

This architecture is particularly promising for ``deep optics''  applications~\cite{wetzstein_inference_2020}, i.e.,~complex processing and recognition of images without converting the optical signal into electronic digital form.
In addition to machine learning tasks, DONNs have further applications in super-resolution imaging and displays, microscopy, quantitative phase imaging, vortex encoding, and coded diffraction imaging~\cite{yan_fourier-space_2019, mengu_all-optical_2022,isil_super-resolution_2022,muminov_fourier_2020, attal_towards_2022}. Several physical realizations of diffractive layers have been experimentally demonstrated using 3D printed materials, metamaterials, and spatial light modulators~\cite{li_spectrally_2021, qian_dynamic_2022, zhou_large-scale_2021}.

DONNs are usually trained \emph{in silico}, i.e.,~the physical DONN is modeled using a computer to simulate the evolution of the optical signals through the system. 
The modulation patterns of the diffractive layers are optimized to achieve the desired transformation between the input and target output of the DONN, which is analogous to optimizing the weights in standard neural network models~\cite{mengu_at_2022}.
During training, the transmittance coefficient of each pixel in the diffractive layers is iteratively updated using an optimization algorithm to minimize the error in the model's output with respect to the training set. 
The backpropagation algorithm is used to efficiently calculate the gradient of the loss with respect to the transmittance coefficients~\cite{rumelhart_learning_1986}.

DONNs are particularly well-suited for use with real-world optical signals, as the optical fields can be directly fed into the system. However, such signals are either spatially incoherent (i.e., the phase of the field at different spatial positions is completely uncorrelated) or only partially coherent (the phase contains some degree of correlation over a limited spatial extent). In contrast, most of the existing experimental work with DONNs is performed using laser sources with fully-coherent illumination (i.e., the phase relationship between all spatial positions is constant).

Optical neural networks with incoherent illumination have been studied by Fei {\it et al.}, albeit in application to convolutional layers~\cite{incoherent_2020}. Rahman {\it et al.} proposed to model incoherent DONNs by averaging the output intensity patterns from numerous coherent input fields with random phase distributions \cite{rahman_universal_2023}.

In this paper, we introduce a computationally efficient framework for simulating and training DONNs using incident illumination  with arbitrary spatial coherence (i.e., fully coherent, partially coherent, or fully incoherent) and discuss the corresponding computational complexity. We also investigate the role of optical coherence on the expressive ability of DONNs to extract complex relationships in the input data. In particular, we show that under fully incoherent illumination, the DONN performance cannot surpass that of a linear classifier. In contrast, when some degree of coherence is present, measuring the intensity at the output constitutes a nonlinear operation, which plays the role of an activation function in standard neural network models. This permits reaching performance levels (e.g.,~classification accuracy) beyond the capabilities of a linear model.
To illustrate our findings, we evaluate the performance of simulated DONNs trained on the MNIST dataset of handwritten digits using light with varying optical coherence and provide the  code in Ref.~\cite{code}. 

\section{DONN Operation with Spatial Coherence}
\subsection{Coherent illumination}

In this section, we introduce a formalism for describing the evolution of coherent, monochromatic optical fields through DONNs using scalar diffraction theory~\cite{goodman_introduction_2017}. 
We treat the optical field as a complex scalar quantity and employ Dirac notation to represent the transverse profile of the field at discrete spatial positions using ket-vectors.
This discretization does not affect the generality of our treatment as long as the spatial sampling interval (i.e., pixel pitch) is much smaller than the characteristic transverse field feature size.
The transformations applied to the field as it evolves through the DONN, which include free-space propagation and transmission through modulation surfaces, are expressed using linear operators.

\begin{table*}[t]
\caption{Computational complexity of modeling the evolution of $B$ input examples through a DONN with $L$ layers under different illumination conditions. Two scenarios are shown: using constant number of pixels ($N$) across all layers and variable number of pixels for each layer, where $N_l$ is the number of pixels at the $l$-th layer and $N_l^* = (N_l + N_{l-1})$.}
\centering
\begin{tabular}{lll}
\toprule
Spatial Coherence & \multicolumn{2}{c}{Computational Complexity} \\
\cmidrule(r){2-3}
         & $N$ Pixels per Layer & Variable Pixels per Layer\\
\midrule
Coherent                  & $\mathcal{O}(BL N\log N)$ & $\sum_{l=1}^{L} \mathcal{O}(B N_l^* \log N_l^*)$ \\
Arbitrary Coherence       & $\mathcal{O}(BL N^2\log N)$ & $\sum_{l=1}^{L} \mathcal{O}(B N_l^{*2}\log N_l^*)$ \\
Incoherent                & $\mathcal{O}(B N^2)+\mathcal{O}(L N^2\log N)$ & $\mathcal{O}(B N_0 N_L)+ \sum_{l=1}^{L} \mathcal{O}(N_l^{*2}\log N_l^{*})$  \\
\bottomrule
\end{tabular}
\label{table:complexity}
\end{table*}

At each layer in the DONN, the optical field is modulated by the diffractive surface and subsequently propagates through free space to the next layer. The incident field at the $m$-th discrete pixel of the $l$-th diffractive layer, before modulation, is represented by $\psi_l(m)$, where the time dependence of the signal is absent.
The transverse profile of the field can be expressed using Dirac notation as $\ket{\psi_l} = \sum_m \psi_l(m) \ket{m}$, where the set  $\{\ket{m}\}$ of all pixels forms an orthonormal basis. The mapping between the optical fields in the $l$-th and ($l$+1)-th layers can be expressed as $\ket{\psi_{l+1}} = \hat{P}_{l} \hat{T}_{l} \ket{\psi_l}$, where $\hat{T}_{l}$ and $\hat{P}_{l}$ are the transmission and free-space propagation operators, respectively.

The transmission operator $\hat{T}_{l}$ describes the phase and/or amplitude modulation applied to the optical field by each pixel in the $l$-th diffractive layer.
The corresponding matrix is diagonal:
\begin{equation}\label{eq:transmission}
    \hat{T}_{l} = \sum_m t_l(m)  \cdot \ketbra{m}{m} ,
\end{equation}
where $t_l(m)$ is the complex-valued transmittance coefficient at the $m$-th pixel in the $l$-th diffractive layer, which satisfies $|t_l(m)| \le 1$.

The operator $\hat{P}_{l}$ describes the free-space propagation of the field between the $l$-th and ($l$+1)-th diffractive layers
using the Rayleigh-Sommerfeld solution~\cite{goodman_introduction_2017}:
\begin{equation}\label{eq:propagation}
    \hat{P}_{l} = \sum_{m,n}  h(m,n) \cdot \ketbra{n}{m} ,
\end{equation}
with
\begin{equation}\label{eq:PSF}
    h(m,n)=\frac{1}{i\lambda} \exp \left(\frac{i2 \pi r(m, n)}{\lambda} \right) \frac{d}{r(m, n)^2} 
\end{equation}
being the point-spread function, i.e.,~the amplitude distribution in the ($l$+1)-th layer if only the $m$-th pixel of the $l$-th layer is illuminated. In the above equation, $\lambda$ is the wavelength of the coherent optical signal (the central wavelength for quasimonochromatic light), $d$ is the axial distance between the two diffractive layers,
and $r(m, n)$ is the Euclidean distance between the $m$-th and $n$-th pixels in the $l$-th and ($l$+1)-th layers, respectively. The above expression is valid when the axial distance between layers is much greater than the (central) wavelength of light.

The input image processed by the DONN is encoded in the initial field $\ket{\psi_\mathrm{in}}$. The output optical field, represented by $\ket{\psi_\mathrm{out}}$, can then be expressed as
\begin{equation}\label{eq:psi_out}
    \ket{\psi_\mathrm{out}} =  \hat{U} \ket{\psi_\mathrm{in}},
\end{equation}
where ${\hat{U}}$ is the evolution operator of the DONN that maps the input optical field onto the output field (i.e., the spatial impulse response of the system):
\begin{equation}\label{eq:U}
    \hat{U} = \prod_{l=1}^L \left(\hat{P}_{l}  \hat{T}_{l} \right) \hat{P}_{0},
\end{equation}
where the DONN has $L$ diffractive layers.
At the output plane of the DONN, the intensity of the evolved field is measured using image sensors:
\begin{equation}\label{eq:I_coherent}
    {I_\mathrm{out}(n)={|\braket{n}{\psi_\mathrm{out}}|^2}}=
|\psi_\mathrm{out}(n)|^2.
\end{equation}

In classification tasks, the output of the DONN for each class $c$ (e.g., the digits from zero to nine in the MNIST handwritten digits dataset) is defined as the total intensity incident on a specified spatial detection region $\mathcal{D}_c$ in the output plane:
\begin{equation}\label{eq:o_n}
    o(c)= \sum_{n\in \mathcal{D}_c}I_\mathrm{out}(n).
\end{equation}

A summary of the mathematical operations applied to the input optical signal by the DONN during inference is shown in Fig.~\ref{fig:diffractive_network_schematic}b.

Training DONNs using a computer (i.e., \textit{in silico}) requires simulating the evolution of coherent optical fields through the system. The calculated optical field at each layer is then used during the backward pass to compute the gradient of the loss function with respect to the diffractive layer transmittance coefficients. 
A naive numerical implementation of the propagation operator $\hat{P}_{l}$ using matrix-vector multiplication has a computational complexity of $\mathcal{O}(N^2)$, where $N$ is the number of pixels per layer. However, this can be optimized by noting that each propagation operator  is described by a Toeplitz matrix since the point-spread function in Eq.~\eqref{eq:PSF} is invariant with respect to translation in space. Hence, the application of this operator constitutes a convolution operation with the input field. 
Thus, the propagation operator $\hat{P}_{l}$ can be evaluated in $\mathcal{O}(N\log N)$ time by utilizing the fast Fourier transform algorithm when both layers contain $N$ pixels~\cite{shen_fast-fourier-transform_2006}.
Additionally, the transmission operator $\hat{T}_{l}$ is described by a diagonal matrix and can be evaluated in $\mathcal{O}(N)$ time. Therefore, calculating the evolution of $B$ different input fields through a DONN with $L$ layers and $N$ pixels per layer has a computational complexity of $\mathcal{O}(BLN\log N)$, and the backward pass has the same complexity. 

\subsection{Arbitrary spatial coherence illumination}
Using DONNs for real-world applications requires the ability to  process incoherent and partially coherent light. 
We assume quasimonochromatic illumination conditions, which is a good approximation for many cases. These conditions require that the input light is narrowband and its coherence length is much greater than the maximum path length difference between diffractive layers~\cite{goodman_statistical_2015}. 
At the same time, we assume the coherence time to be much shorter than the inverse detection bandwidth, so the detection averages in time over the non-stationary interference pattern.

The spatial coherence of the optical field in the $l$-th layer is characterized by the mutual intensity function, which determines the time-averaged correlation of the field at two separate pixels~\cite{mandel_optical_1995,goodman_statistical_2015}:
$$J_l(m,m')=\lim_{T\rightarrow\infty} \frac{1}{T} \int_{-T/2}^{T/2} \psi_l(m; t) \, \psi^*_l(m';t) \, \mathrm{d}t,$$
where $T$ is the detection time. This matrix represents an operator
\begin{equation}
    \hat{J}_{l} = \lim_{T\rightarrow\infty} \frac{1}{T} \int_{-T/2}^{T/2} \ketbra{\psi_l(t)}{\psi_l(t)}\,\mathrm{d}t,
\end{equation}
where $J_l(m,m')=\mel{m}{\hat{J}_{l}}{m'}$. The time-averaged intensity of the field is given by the diagonal elements of the mutual intensity operator, such that 
\begin{equation}\label{Intinc}
    I_l(m)=J_{l}(m,m).
\end{equation}

Similar to the evolution of coherent fields through DONNs, the evolution of the mutual intensity operator can be expressed using the transmission and propagation operators. 
The input mutual intensity operator $\hat{J}_{\mathrm{in}}$ describes the spatial coherence of the initial field that encodes the input image to be processed by the DONN. The output mutual intensity operator is given by
\begin{equation}\label{eq:J_out}
    {\hat{J}_{\mathrm{out}} = \hat{U} \, \hat{J}_{\mathrm{in}} \, \hat{U}^\dagger},
\end{equation}
where $\hat{U}$ is the evolution operator of the DONN defined in Eq.~\eqref{eq:U}. Analogous to the coherent case, the output of the DONN corresponding to each class $c$ is the total time-averaged intensity, defined in Eq.~\eqref{Intinc}, incident on the spatial detection regions along the output plane:
\begin{equation}\label{eq:o_n_coherence}
    o(c)= \sum_{n \in \mathcal{D}_c}{J}_{\mathrm{out}}(n,n) = \sum_{n \in \mathcal{D}_c}{I}_{\mathrm{out}}(n).
\end{equation}

The evolution of the mutual intensity operator and the corresponding DONN output can be simulated on a computer using Eqs.~\eqref{eq:J_out} and~\eqref{eq:o_n_coherence}. Analogous to the previously discussed method, the fast Fourier transform can be leveraged to evaluate the propagation operator $\hat{P}_l$ applied to an arbitrary mutual intensity operator described by an $N\times N$ matrix, which scales as $\mathcal{O}(N^2\log N)$. The transmission operator $\hat{T}_l$ can similarly be evaluated in $\mathcal{O}(N^2)$ time. Hence, simulating the evolution of $B$ different input fields with arbitrary spatial coherence through a DONN with $L$ layers and $N$ pixels per layer has a computational complexity of $\mathcal{O}(B L N^2 \log N)$. The backward pass executed during training has the same computational complexity. 

\subsection{Incoherent illumination}
DONNs with fully incoherent input can be treated using the computational approach discussed in the previous section. However, computational costs can be amortised for multiple input examples using the impulse response that characterizes the system.  

Fully incoherent input is described by the diagonal mutual intensity operator
\begin{equation}\label{Jinc}
    J_\mathrm{in}(m,m')=I_\mathrm{in}(m) \, \delta_{m,m'},
\end{equation}
where the time-averaged intensity $I_\mathrm{in}(m)$ encodes information from the $m$-th pixel of the input image. We can express the corresponding time-averaged intensity along the output plane, using Eqs.~\eqref{Intinc} and \eqref{eq:J_out}, as
\begin{equation}\label{eq:I_incoherent}
    I_\mathrm{out}(n) = \sum_m I_\mathrm{in}(m) \cdot \left|\mel{n}{\hat{U}}{m} \right|^2.
\end{equation}
Here, $|\mel{n}{\hat{U}}{m}|^2$ corresponds to the intensity at the $n$-th pixel in the output plane 
from point-source illumination at the $m$-th input pixel (i.e., the intensity impulse response of the system).

\begin{algorithm}[H]
\caption{Training DONNs with spatial coherence}
\label{alg:train}
\begin{algorithmic}[1]
\State \textbf{Input:} Set of input examples $X$ (field $\psi$ if coherent, intensity $I$ if incoherent, otherwise mutual intensity $J$), corresponding labels $Y$, loss function $\mathcal{L}$, initial DONN parameters $\theta$ (transmittance coefficients $t_l$ in each layer), number of epochs $E$, and the optimizer.
\For {$epoch=1$  to $E$}
\For{each $(x,y)$ pair in ($X, Y$)}
    \State Compute output intensity $I_\mathrm{out}$ \Comment{Eq. \eqref{eq:I_coherent}, \eqref{eq:J_out},  \eqref{eq:I_incoherent}}
    \State Calculate DONN output $o$  \Comment{Eq. \eqref{eq:o_n_coherence}}
    \State Determine Loss $\mathcal{L}(y,o)$
    \State Compute $\nabla_\theta \mathcal{L}$ using backpropagation
    \State Update $\theta$ using the optimizer 
\EndFor
\EndFor
\State \textbf{Output:} Optimized DONN parameters $\theta$
\end{algorithmic}
\end{algorithm}

The intensity impulse response of a DONN with $L$ layers and $N$ pixels per layer can be determined by calculating the coherent evolution of all $N$ input pixels through the system, which
has a computational complexity of $\mathcal{O}(LN^2 \log N)$. This is the same complexity as the previous method using the evolution of the mutual intensity operator. 
However, once the intensity impulse response is known, the output intensity distributions for $B$ different incoherent input fields can be calculated using Eq.~\eqref{eq:I_incoherent} with a reduced computational complexity of $\mathcal{O}(BN^2)$.
Thus, the total computational cost of simulating DONN inference with fully incoherent input for multiple input examples can be reduced.
This technique can be implemented during training to improve the simulation runtime by using mini-batches of input examples, as calculating the unit responses of the DONN is only required once for each mini-batch.

A deep-learning based method  was recently proposed for implementing linear transformations with DONNs under spatially incoherent illumination~\cite{rahman_universal_2023}. 
The method approximates incoherence for a single input example by averaging the output intensity patterns from numerous coherent input fields with random phase distributions.
For example, in Ref.~\cite{rahman_universal_2023}, the authors use 20,000 random phase patterns during the testing phase to compute the incoherent output intensity for a single $8\times8$ input example.
In contrast, our approach calculates the exact output intensity using the intensity impulse response of the system, which only requires $64$ coherent input fields. 

A summary of the computational complexities of simulating DONNs under coherent, arbitrary coherence, and incoherent illumination is shown in Table~\ref{table:complexity}. The general training procedure is given in Algorithm~\ref{alg:train}, where the DONN output $o$ is computed using the previously described methods.

\begin{figure*}[t]
    \centering
    \includegraphics[width=0.99\linewidth]{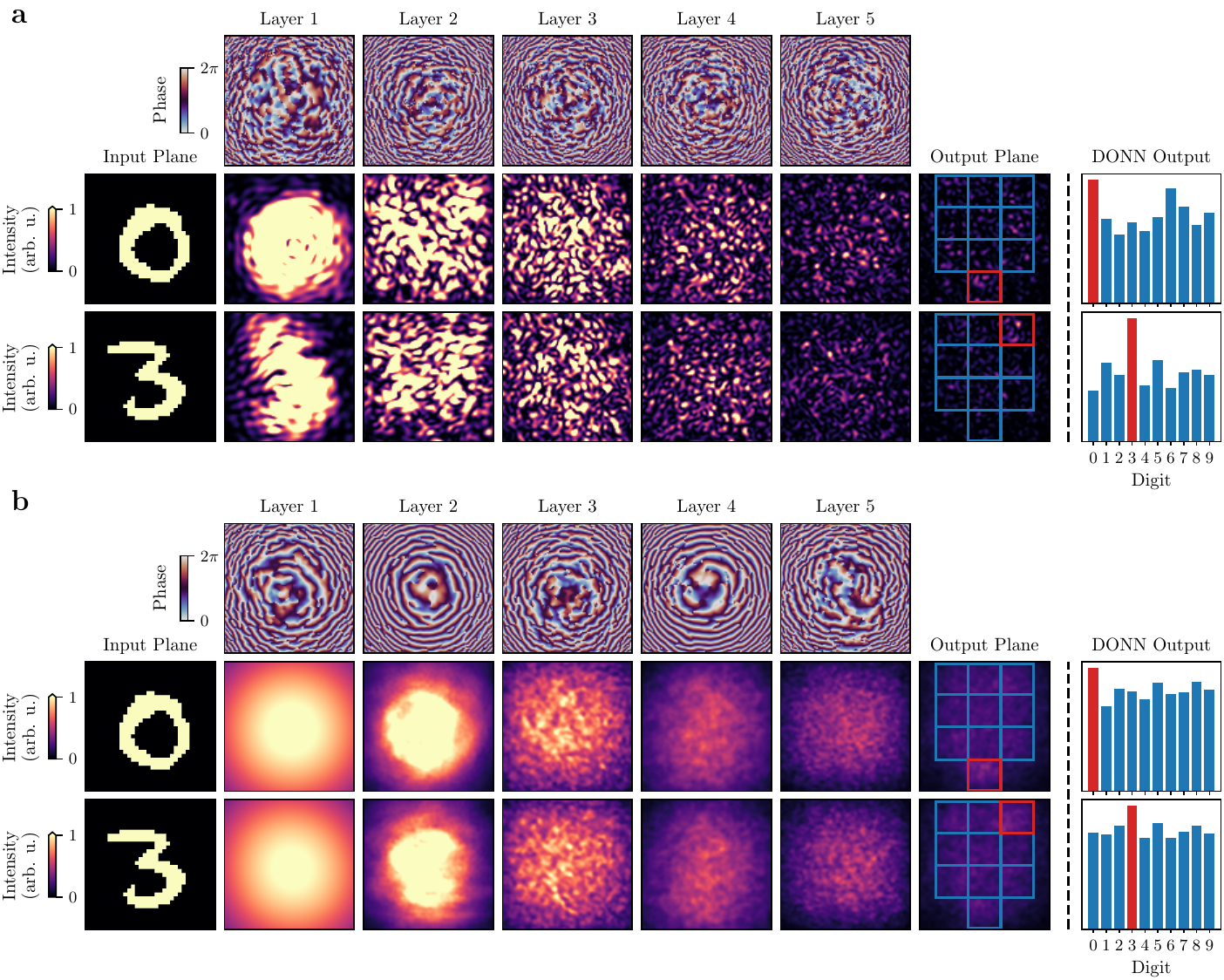}
    \caption{Evolution of input examples in a DONN with five layers trained using coherent and incoherent illumination. \textbf{a} The top row shows the phase-only modulation profiles of the five diffractive layers, which were trained to process the MNIST dataset using coherent illumination. In the middle row, the digit zero is fed into the system in the input layer and the time-averaged intensity at each layer is shown. The detection regions at the output layer are indicated using blue and red boxes, where the red box corresponds to the target region. The output of the DONN, which is the total intensity in each detection region, is shown on the right. In the bottom row, the evolution of the digit three is shown. The same scaling for the intensity values is used across each row. \textbf{b}~The equivalent visualization of a DONN trained using incoherent illumination.}
    \label{fig:transmission_layers}
\end{figure*}

\subsection{Expressivity of DONNs}
The expressive power of DONNs is dependent on the spatial coherence of the input light. Under coherent illumination, DONNs have been shown to outperform linear models~\cite{lin_all-optical_2018,spall_hybrid_2022}. Since the coherent field evolves linearly through the system, this improvement in performance results from the nonlinear intensity measurement of the complex-valued field at the output plane ~\eqref{eq:I_coherent}, followed by the linear summation of the intensities over the detection regions ~\eqref{eq:o_n}. Thus, DONNs using coherent illumination can be understood as standard neural networks that consist of a complex-valued linear layer with a nonlinear activation function, followed by a real-valued linear layer, as shown in Fig.~\ref{fig:diffractive_network_schematic}b.

\begin{figure*}[t]
    \centering
    \includegraphics[width=0.8\linewidth]{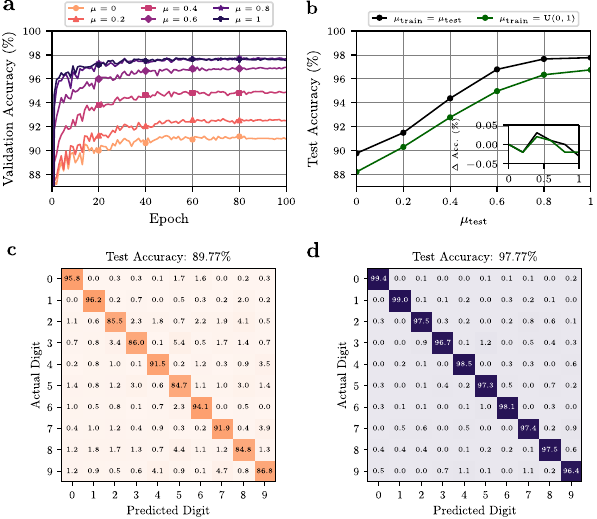}
    \caption{DONN training results on the MNIST dataset.
        \textbf{a}~Validation accuracy at each epoch during training.
        \textbf{b}~Test accuracy using input light with a degree of spatial coherence $\mu_\mathrm{test}$. The black curve corresponds to the performance of models trained to process optical signals with a specified degree of spatial coherence ($\mu_\mathrm{train} = \mu_\mathrm{test}$). The green curve corresponds to the test accuracy achieved by a model trained to process optical signals with any degree of spatial coherence ($\mu_\mathrm{train} = U(0,1)$). The inset plot shows the change in test accuracy  using transmittance coefficients in the diffractive layers with 8-bit precision.
        \textbf{c, d}~Confusion matrices for the incoherent and coherent models, respectively, on the test set.
        }
        \label{fig:Fig3}
\end{figure*}
In contrast, under incoherent illumination, the time-averaged output intensity is the sum of the intensity patterns from individual pixel sources in the input plane. Therefore, DONNs with incoherent input illumination cannot perform better than a linear model, as the time-averaged input and output intensity distributions are linearly related according to the intensity impulse response, as shown in Eq.~\eqref{eq:I_incoherent}.

The improved expressive power of a coherently illuminated DONN arises from the off-diagonal elements in the input mutual intensity operator, which are absent for incoherent light. These elements represent the spatial coherence between two different pixels in the input image. For an arbitrary input mutual intensity operator $J_\mathrm{in}$, the output intensity can be expressed, using Eqs.~\eqref{Intinc} and \eqref{eq:J_out}, as 
\begin{equation}\label{I_sec3}
    I_\mathrm{out}(n) = \sum_{m, m'} J_\mathrm{in} (m, m') \cdot \mel{n}{\hat{U}}{m} \cdot \mel{m'}{\hat{U}^\dagger}{n}.
\end{equation}
This summation includes off-diagonal elements of the mutual intensity operator, which depend nonlinearly on the input field. 
Due to this nonlinearity, the performance of DONNs under partially coherent illumination can surpass that with incoherent light, as demonstrated in the following section. 

\section{Performance on MNIST dataset}
We trained simulated DONN models to identify handwritten digits from zero to nine using incoherent, partially coherent, and coherent illumination.
The models were trained over 100 epochs using 55,000 images (plus 5,000 images for validation) from the MNIST dataset, each consisting of $28\times28$ pixels~\cite{deng_mnist_2012}.
The DONNs are composed of five diffractive layers, each with $100\times100$ pixels, which modulate the phase of the incident light and are spaced 2~cm apart.
Each model was trained using a uniform, normalized optical field incident on the input image of the handwritten digit with a central wavelength of 700~nm.
The cross-entropy loss function was used during training to calculate the output error of the model.
Each pixel in the diffractive layers has a surface area of $10 \times 10\  \si{\micro\meter^2}$, while each pixel in the input pattern is $40 \times 40\ \si{\micro\meter^2}$.
Each detection region in the output plane, which is associated with a unique digit, has an area of $250 \times 250\ \si{\micro\meter^2}$. DONNs trained under incoherent and coherent illumination are illustrated in Fig.~\ref{fig:transmission_layers}.

For each image in the dataset, we define the spatial coherence of the input illumination as
\begin{equation}
    J_\mathrm{in}(m, m') = \sqrt{I_\mathrm{im}(m) I_\mathrm{im}(m')} \, \mu^{{r_\mathrm{norm}}(m, m')},
\end{equation}
where $I_\mathrm{im}(m)$ is the time-averaged intensity at the $m$-th pixel in the image, $r_\mathrm{norm}(m, m')$ is the Euclidean distance between the $m$-th and $m'$-th input pixels, normalized by the pixel pitch, and $\mu$ quantifies the degree of spatial coherence: $\mu=0$ for fully incoherent  and $\mu=1$ for fully coherent light. We compute the intensity at the output plane using Eq.~\eqref{I_sec3}, which is more computationally efficient than calculating the evolution of the mutual intensity operator at each layer since the input images contain fewer pixels than the diffractive layers.

We first trained two DONNs to process handwritten digits using fully incoherent and coherent illumination.
During the training phase, we used a learning rate of 0.02 with exponential decay factor of $\gamma=0.95$ and saved the model parameters that yielded the highest validation accuracy.  We then evaluated the performance of these models using a test set of 10,000 images that were not shown during training. 
The models trained using incoherent and coherent light achieved test accuracies of  89.77\% and 97.77\%, respectively. For comparison, we trained a standard linear classifier model and achieved a test accuracy of 92.57\%, which outperforms the incoherent DONN as expected.
We then trained DONNs using partially coherent input illumination where each model was trained to process light with a different degree of spatial coherence ($\mu=0.2,0.4,0.6,0.8$). 
The validation accuracy attained by these models during training, as well as their performance on the test set and the confusion matrices for incoherent and coherent light, are shown in Fig.~\ref{fig:Fig3}. The optimal performance is achieved using coherent light. 
The models were also evaluated using transmittance coefficients in the diffractive layers with 8-bit precision to replicate experimental implementations (e.g., spatial light modulators); however, the change in performance is negligible.

Finally, we trained a DONN to process optical signals with variable degree of spatial coherence~$\mu$. During the training phase, we randomly varied $\mu$ between 0 and 1 for each input example. We evaluated the model on the test set across different degrees of spatial coherence $\mu_\mathrm{test}$, and the results are illustrated in Fig.~\ref{fig:Fig3}b.
The model demonstrated a high level of accuracy, achieving 
88.21\% for incoherent light ($\mu_\mathrm{test}=0$) and 96.75\% for coherent light ($\mu_\mathrm{test}=1$)  . This corresponds to a decrease in accuracy of $<2\%$ when compared to DONNs tailored to process a specific $\mu$ value.

\section{Conclusion}

We have demonstrated that the performance of DONNs is dependent on the spatial coherence of the incident illumination. Models using incoherent illumination cannot outperform linear models for information processing tasks.
However, as demonstrated in Fig.~\ref{fig:Fig3}b, the degree of spatial coherence required to achieve optimal performance need not be high: $\mu\sim 0.6$ means that the mutual coherence between points separated by four pixels is reduced by a factor of $0.13$. That is, performance almost at the fully coherent level can be reached even when the transverse coherence length is much less than the size of a whole MNIST digit.
This implies that neighboring pixels contain more relevant information for pattern recognition compared to distant pixels.
As a result, the DONN model can capture relevant nonlinear relationships in the input data without requiring full spatial coherence between all pixels.

We emphasize that the above relation between the input coherence and DONN expressivity assumes that no further processing of the DONN data is implemented. If, for example, the DONN is followed by an electronic neural network with nonlinear activation layers, the DONN can surpass a linear model even if illuminated incoherently. For example, Rahman {\it et al.} trained a DONN to classify MNIST by associating two detection regions with each digit and then applying a rational function to compute  the network prediction from the intensities of these regions. In this way, the accuracy reached was above that of a linear classifier~\cite{rahman_universal_2023}.

Incoherently illuminated DONNs are more broadly applicable to real-world environments, as coherent illumination requires a laser source. However, some degree of coherence can also be achieved by illuminating the object with a distant incoherent source of narrow spatial extent according to the van Cittert --- Zernike theorem \cite{mandel_optical_1995}. As discussed above, illumination with even a short transverse coherence length can significantly enhance the DONN performance.

In addition to the coherence of the incident light,  DONN performance depends on several system properties including the number of layers, the number of pixels in each layer, the geometry of the detection regions, the optical wavelength, the  axial distances between layers, and the pixel size. Similar to standard digital neural networks, increasing the number of trainable parameters, including the number of layers, improves the overall expressibility of the system. The geometric properties of the DONN can be encapsulated in the Fresnel number for each pair of subsequent layers, which linearly relates the wavelength, propagation distance, and layer size~\cite{goodman_introduction_2017}. These numbers can be optimized as learnable parameters during training, where the optimal values depend on the training dataset. Based on the constraints of the experimental setup and the optimized Fresnel numbers, the aperture size, wavelength, and propagation distances can be selected. Finally, it is important to ensure that the pixel size in each layer is significantly smaller than the features in the layer's optimized diffraction pattern. 

The robustness of the system can be improved by training the DONN using various illumination conditions, which could be useful for applications that require DONN operation in different environments. Moreover, the system can be further generalized to operate under a continuum of central frequencies, which has been recently experimentally demonstrated using coherent light~\cite{luo_design_2019}.

\section*{Acknowledgments}
This  work is supported by Innovate UK Smart Grant 10043476. M.J.F. is funded by the European Union’s Horizon 2020 research and innovation programme under the Marie Skłodowska-Curie grant agreement No.~956071.


\begin{thebibliography}{99}

\bibitem{Goodfellow-et-al-2016}
I.~Goodfellow, Y.~Bengio, and A.~Courville, {\em Deep Learning}.
\newblock MIT Press, 2016.

\bibitem{bochkovskiy_yolov4_2020}
A.~Bochkovskiy, C.-Y. Wang, and H.-Y.~M. Liao, ``{YOLOv4}: {Optimal} {Speed} and {Accuracy} of {Object} {Detection},'' Apr. 2020.
\newblock arXiv:2004.10934 [cs, eess].

\bibitem{brown_language_2020}
T.~Brown, B.~Mann, N.~Ryder, M.~Subbiah, J.~D. Kaplan, P.~Dhariwal, A.~Neelakantan, P.~Shyam, G.~Sastry, A.~Askell, S.~Agarwal, A.~Herbert-Voss, G.~Krueger, T.~Henighan, R.~Child, A.~Ramesh, D.~Ziegler, J.~Wu, C.~Winter, C.~Hesse, M.~Chen, E.~Sigler, M.~Litwin, S.~Gray, B.~Chess, J.~Clark, C.~Berner, S.~McCandlish, A.~Radford, I.~Sutskever, and D.~Amodei, ``Language {Models} are {Few}-{Shot} {Learners},'' in {\em Advances in {Neural} {Information} {Processing} {Systems}}, vol.~33, pp.~1877--1901, Curran Associates, Inc., 2020.

\bibitem{senior_improved_2020}
A.~W. Senior, R.~Evans, J.~Jumper, J.~Kirkpatrick, L.~Sifre, T.~Green, C.~Qin, A.~Žídek, A.~W.~R. Nelson, A.~Bridgland, H.~Penedones, S.~Petersen, K.~Simonyan, S.~Crossan, P.~Kohli, D.~T. Jones, D.~Silver, K.~Kavukcuoglu, and D.~Hassabis, ``Improved protein structure prediction using potentials from deep learning,'' {\em Nature}, vol.~577, pp.~706--710, Jan. 2020.

\bibitem{dario_amodei_ai_2018}
{Dario Amodei} and {Danny Hernandez}, ``{AI} and {Compute},'' May 2018.

\bibitem{thompson_computational_2022}
N.~C. Thompson, K.~Greenewald, K.~Lee, and G.~F. Manso, ``The {Computational} {Limits} of {Deep} {Learning},'' July 2022.
\newblock arXiv:2007.05558 [cs, stat].

\bibitem{mehonic_brain-inspired_2022}
A.~Mehonic and A.~J. Kenyon, ``Brain-inspired computing needs a master plan,'' {\em Nature}, vol.~604, pp.~255--260, Apr. 2022.

\bibitem{shastri_photonics_2021}
B.~J. Shastri, A.~N. Tait, T.~Ferreira~de Lima, W.~H.~P. Pernice, H.~Bhaskaran, C.~D. Wright, and P.~R. Prucnal, ``Photonics for artificial intelligence and neuromorphic computing,'' {\em Nature Photonics}, vol.~15, pp.~102--114, Feb. 2021.

\bibitem{vuong_what_2023}
L.~T. Vuong, D.~G. Stavenga, and G.~L. Barrows, ``What computer vision can learn from insect vision,'' {\em Optics \& Photonics News}, vol.~34, pp.~24, Nov. 2023.

\bibitem{wang_image_2023}
T.~Wang, M.~M. Sohoni, L.~G. Wright, M.~M. Stein, S.-Y. Ma, T.~Onodera, M.~G. Anderson, and P.~L. {McMahon}, ``Image sensing with multilayer nonlinear optical neural networks,'' {\em Nature Photonics}, pp.~1--8, 2023.

\bibitem{spall_fully_2020}
J.~Spall, X.~Guo, T.~D. Barrett, and A.~I. Lvovsky, ``Fully reconfigurable coherent optical vector-matrix multiplication,'' {\em Optics Letters}, vol.~45, no.~20, p.~5752, 2020.

\bibitem{miscuglio_massively_2020}
M.~Miscuglio, Z.~Hu, S.~Li, J.~K. George, R.~Capanna, H.~Dalir, P.~M. Bardet, P.~Gupta, and V.~J. Sorger, ``Massively parallel amplitude-only {Fourier} neural network,'' {\em Optica}, vol.~7, no.~12, pp.~1812--1819, 2020.

\bibitem{zuo_all-optical_2019}
Y.~Zuo, B.~Li, Y.~Zhao, Y.~Jiang, Y.-C. Chen, P.~Chen, G.-B. Jo, J.~Liu, and S.~Du, ``All-optical neural network with nonlinear activation functions,'' {\em Optica}, vol.~6, no.~9, pp.~1132--1137, 2019.

\bibitem{zhang_silicon_2022}
W.~Zhang, C.~Huang, H.-T. Peng, S.~Bilodeau, A.~Jha, E.~Blow, T.~F. de~Lima, B.~J. Shastri, and P.~Prucnal, ``Silicon microring synapses enable photonic deep learning beyond 9-bit precision,'' {\em Optica}, vol.~9, p.~579, May 2022.

\bibitem{ashtiani_-chip_2022}
F.~Ashtiani, A.~J. Geers, and F.~Aflatouni, ``An on-chip photonic deep neural network for image classification,'' {\em Nature}, vol.~606, pp.~501--506, June 2022.

\bibitem{feldmann_all-optical_2019}
J.~Feldmann, N.~Youngblood, C.~D. Wright, H.~Bhaskaran, and W.~H.~P. Pernice, ``All-optical spiking neurosynaptic networks with self-learning capabilities,'' {\em Nature}, vol.~569, no.~7755, p.~208, 2019.

\bibitem{tait_neuromorphic_2017}
A.~N. Tait, T.~F.~d. Lima, E.~Zhou, A.~X. Wu, M.~A. Nahmias, B.~J. Shastri, and P.~R. Prucnal, ``Neuromorphic photonic networks using silicon photonic weight banks,'' {\em Scientific Reports}, vol.~7, no.~1, p.~7430, 2017.

\bibitem{xu_11_2021}
X.~Xu, M.~Tan, B.~Corcoran, J.~Wu, A.~Boes, T.~G. Nguyen, S.~T. Chu, B.~E. Little, D.~G. Hicks, R.~Morandotti, A.~Mitchell, and D.~J. Moss, ``11 {TOPS} photonic convolutional accelerator for optical neural networks,'' {\em Nature}, vol.~589, no.~7840, pp.~44--51, 2021.

\bibitem{feldmann_parallel_2021}
J.~Feldmann, N.~Youngblood, M.~Karpov, H.~Gehring, X.~Li, M.~Stappers, M.~Le~Gallo, X.~Fu, A.~Lukashchuk, A.~S. Raja, J.~Liu, C.~D. Wright, A.~Sebastian, T.~J. Kippenberg, W.~H.~P. Pernice, and H.~Bhaskaran, ``Parallel convolutional processing using an integrated photonic tensor core,'' {\em Nature}, vol.~589, no.~7840, pp.~52--58, 2021.

\bibitem{ma_quantum-noise-limited_2023}
S.-Y. Ma, T.~Wang, J.~Laydevant, L.~G. Wright, and P.~L. McMahon, ``Quantum-noise-limited optical neural networks operating at a few quanta per activation,'' July 2023.
\newblock arXiv:2307.15712 [physics, physics:quant-ph].

\bibitem{wang_optical_2022}
T.~Wang, S.-Y. Ma, L.~G. Wright, T.~Onodera, B.~C. Richard, and P.~L. {McMahon}, ``An optical neural network using less than 1 photon per multiplication,'' {\em Nature Communications}, vol.~13, no.~1, p.~123, 2022.

\bibitem{spall_training_2023}
J.~Spall, X.~Guo, and A.~I. Lvovsky, ``Training neural networks with end-to-end optical backpropagation,'' Aug. 2023.
\newblock arXiv:2308.05226 [physics].

\bibitem{pai_experimentally_2023}
S.~Pai, Z.~Sun, T.~W. Hughes, T.~Park, B.~Bartlett, I.~A.~D. Williamson, M.~Minkov, M.~Milanizadeh, N.~Abebe, F.~Morichetti, A.~Melloni, S.~Fan, O.~Solgaard, and D.~A.~B. Miller, ``Experimentally realized in situ backpropagation for deep learning in photonic neural networks,'' {\em Science}, vol.~380, pp.~398--404, Apr. 2023.

\bibitem{filipovich_silicon_2022}
M.~J. Filipovich, Z.~Guo, M.~Al-Qadasi, B.~A. Marquez, H.~D. Morison, V.~J. Sorger, P.~R. Prucnal, S.~Shekhar, and B.~J. Shastri, ``Silicon photonic architecture for training deep neural networks with direct feedback alignment,'' {\em Optica}, vol.~9, no.~12, pp.~1323--1332, 2022.

\bibitem{spall_hybrid_2022}
J.~Spall, X.~Guo, and A.~I. Lvovsky, ``Hybrid training of optical neural networks,'' {\em Optica}, vol.~9, no.~7, p.~803, 2022.

\bibitem{zhou_situ_2020}
T.~Zhou, L.~Fang, T.~Yan, J.~Wu, Y.~Li, J.~Fan, H.~Wu, X.~Lin, and Q.~Dai, ``In situ optical backpropagation training of diffractive optical neural networks,'' {\em Photonics Research}, vol.~8, no.~6, p.~940, 2020.

\bibitem{hughes_training_2018}
T.~W. Hughes, M.~Minkov, Y.~Shi, and S.~Fan, ``Training of photonic neural networks through in situ backpropagation and gradient measurement,'' {\em Optica}, vol.~5, no.~7, p.~864, 2018.

\bibitem{lin_all-optical_2018}
X.~Lin, Y.~Rivenson, N.~T. Yardimci, M.~Veli, Y.~Luo, M.~Jarrahi, and A.~Ozcan, ``All-optical machine learning using diffractive deep neural networks,'' {\em Science}, vol.~361, pp.~1004--1008, Sept. 2018.

\bibitem{mengu_analysis_2020}
D.~Mengu, Y.~Luo, Y.~Rivenson, and A.~Ozcan, ``Analysis of diffractive optical neural networks and their integration with electronic neural networks,'' {\em {IEEE} Journal of Selected Topics in Quantum Electronics}, vol.~26, no.~1, pp.~1--14, 2020.

\bibitem{incoherent_2020}
Y. Fei, X. Sui, and G. Gu and Q. Chen, ``Zero-power optical convolutional neural network using incoherent light,`` {\em Optics and Lasers in Engineering}, vol.~162, pp.~107410, 2023.

\bibitem{wetzstein_inference_2020}
G.~Wetzstein, A.~Ozcan, S.~Gigan, S.~Fan, D.~Englund, M.~Soljačić, C.~Denz, D.~A.~B. Miller, and D.~Psaltis, ``Inference in artificial intelligence with deep optics and photonics,'' {\em Nature}, vol.~588, pp.~39--47, Dec. 2020.

\bibitem{yan_fourier-space_2019}
T.~Yan, J.~Wu, T.~Zhou, H.~Xie, F.~Xu, J.~Fan, L.~Fang, X.~Lin, and Q.~Dai, ``Fourier-space diffractive deep neural network,'' {\em Physical Review Letters}, vol.~123, no.~2, p.~023901, 2019.

\bibitem{mengu_all-optical_2022}
D.~Mengu and A.~Ozcan, ``All-optical phase recovery: Diffractive computing for quantitative phase imaging,'' {\em Advanced Optical Materials}, vol.~10, no.~15, p.~2200281, 2022.

\bibitem{isil_super-resolution_2022}
C.~Isil, D.~Mengu, Y.~Zhao, A.~Tabassum, J.~Li, Y.~Luo, M.~Jarrahi, and A.~Ozcan, ``Super-resolution image display using diffractive decoders,'' {\em Science Advances}, vol.~8, no.~48, p.~eadd3433, 2022.

\bibitem{muminov_fourier_2020}
B.~Muminov and L.~T. Vuong, ``Fourier optical preprocessing in lieu of deep learning,'' {\em Optica}, vol.~7, no.~9, p.~1079, 2020.

\bibitem{attal_towards_2022}
B.~Attal and M.~O'Toole, ``Towards Mixed-State Coded Diffraction Imaging,'' {\em IEEE Trans. Pattern Anal. Mach. Intell.}, p.~1--12, 2022.

\bibitem{li_spectrally_2021}
J.~Li, D.~Mengu, N.~T. Yardimci, Y.~Luo, X.~Li, M.~Veli, Y.~Rivenson, M.~Jarrahi, and A.~Ozcan, ``Spectrally encoded single-pixel machine vision using diffractive networks,'' {\em Science Advances}, vol.~7, no.~13, p.~769, 2021.

\bibitem{qian_dynamic_2022}
C.~Qian, Z.~Wang, H.~Qian, T.~Cai, B.~Zheng, X.~Lin, Y.~Shen, I.~Kaminer, E.~Li, and H.~Chen, ``Dynamic recognition and mirage using neuro-metamaterials,'' {\em Nature Communications}, vol.~13, no.~1, p.~2694, 2022.

\bibitem{zhou_large-scale_2021}
T.~Zhou, X.~Lin, J.~Wu, Y.~Chen, H.~Xie, Y.~Li, J.~Fan, H.~Wu, L.~Fang, and Q.~Dai, ``Large-scale neuromorphic optoelectronic computing with a reconfigurable diffractive processing unit,'' {\em Nature Photonics}, vol.~15, no.~5, pp.~367--373, 2021.

\bibitem{mengu_at_2022}
D.~Mengu, M.~S. Sakib~Rahman, Y.~Luo, J.~Li, O.~Kulce, and A.~Ozcan, ``At the intersection of optics and deep learning: statistical inference, computing, and inverse design,'' {\em Advances in Optics and Photonics}, vol.~14, no.~2, p.~209, 2022.

\bibitem{rumelhart_learning_1986}
D.~E. Rumelhart, G.~E. Hinton, and R.~J. Williams, ``Learning representations by back-propagating errors,'' {\em Nature}, vol.~323, pp.~533--536, Oct. 1986.

\bibitem{code} GitHub repository:
\href{https://github.com/MatthewFilipovich/diffractive-optical-neural-networks-with-coherence}{https://github.com/MatthewFilipovich /diffractive-optical-neural-networks-with-coherence}

\bibitem{rahman_universal_2023}
M.~S.~S. Rahman, X.~Yang, J.~Li, B.~Bai, and A.~Ozcan, ``Universal linear intensity transformations using spatially incoherent diffractive processors,'' {\em Light: Science \& Applications}, vol.~12, p.~195, Aug. 2023.

\bibitem{mandel_optical_1995}
L.~Mandel and E.~Wolf, {\em Optical {Coherence} and {Quantum} {Optics}}.
\newblock Cambridge: Cambridge University Press, 1995.

\bibitem{goodman_introduction_2017}
J.~W. Goodman, {\em Introduction to {Fourier} optics}.
\newblock W. H. Freeman, 2017.

\bibitem{shen_fast-fourier-transform_2006}
F.~Shen and A.~Wang, ``Fast-{Fourier}-transform based numerical integration method for the {Rayleigh}-{Sommerfeld} diffraction formula,'' {\em Applied Optics}, vol.~45, p.~1102, Feb. 2006.

\bibitem{goodman_statistical_2015}
J.~W. Goodman, {\em Statistical {Optics}}.
\newblock John Wiley \& Sons, May 2015.

\bibitem{deng_mnist_2012}
L.~Deng, ``The {MNIST} database of handwritten digit images for machine learning research,'' {\em {IEEE} Signal Processing Magazine}, vol.~29, no.~6, pp.~141--142, 2012.

\bibitem{luo_design_2019}
Y.~Luo, D.~Mengu, N.~T. Yardimci, Y.~Rivenson, M.~Veli, M.~Jarrahi, and A.~Ozcan, ``Design of task-specific optical systems using broadband diffractive neural networks,'' {\em Light: Science \& Applications}, vol.~8, no.~1, p.~112, 2019.

\end{thebibliography}
\end{document}